\documentclass[a4paper,12pt]{article}
\title{Finite N Index and Angular Momentum Bound from Gravity}
\author{Yu Nakayama}
\usepackage{amsmath}
\usepackage{amssymb}
\usepackage{graphicx}
\usepackage{hyperref}
\def\drawbox#1#2{\hrule height#2pt
        \hbox{\vrule width#2pt height#1pt \kern#1pt
              \vrule width#2pt}
              \hrule height#2pt}

\def\Fund#1#2{\vcenter{\vbox{\drawbox{#1}{#2}}}}
\def\Asym#1#2{\vcenter{\vbox{\drawbox{#1}{#2}
              \kern-#2pt       
              \drawbox{#1}{#2}}}}

\def\funda{\Fund{6.5}{0.4}}

\def\symm{\funda\kern-0.4pt\funda}

\allowdisplaybreaks[0]

\newcommand{\sectiono}[1]{\section{#1}\setcounter{equation}{0}}

\begin{document}

\begin{titlepage}
\thispagestyle{empty}
\begin{flushright}
UT-07-04\\
hep-th/yymmnnn\\
\end{flushright}

\vskip 1.5 cm

\begin{center}
\noindent{\textbf{\LARGE{\\\vspace{0.5cm} Finite $N$ Index}}}
{\textbf{\LARGE{\\\vspace{0.5cm} and Angular Momentum Bound from Gravity}}}
\vskip 1.5cm
\noindent{\large{Yu Nakayama}\footnote{E-mail: nakayama@hep-th.phys.s.u-tokyo.ac.jp}}\\ 
\vspace{1cm}
\noindent{\small{\textit{Department of Physics, Faculty of Science, University of 
Tokyo}} \\ \vspace{2mm}
\small{\textit{Hongo 7-3-1, Bunkyo-ku, Tokyo 113-0033, Japan}}}
\end{center}
\vspace{1cm}
\begin{abstract}
We exactly compute the finite $N$ index and BPS partition functions for $\mathcal{N}=4$ SYM theory in a newly proposed maximal angular momentum limit. The new limit is not predicted from the superconformal algebra, but naturally arises from the supergravity dual. We show that the index does not receive any finite $N$ corrections while the free BPS partition function does.

\end{abstract}

\end{titlepage}


\sectiono{Introduction}\label{sec:1}

Finite $N$ corrections to the super Yang-Mills (SYM) theory are of great significance in the contex of the AdS-CFT correspondence both technically and conceptually. Since $N$ in the gauge theory is naturally quantized, the spectrum of the type IIB string theory in the AdS space should also be quantized nonperturbatively beyond the supergravity approximation. In this sense, any enumeration programs of the physical states in the finite $N$ theory, or finite string coupling constant would yield an important clue to the nonperturbative definition of the string theory and its basic ingredient.

One of the most fundamental enumeration programs for this purpose is the index of superconformal field theories (SCFTs) \cite{Romelsberger:2005eg,Kinney:2005ej}. The index is the unique quantity invariant under any marginal deformations of the SCFTs and plays a central role in classifying the SCFTs \cite{Nakayama:2005mf,Nakayama:2006ur}. Moreover it was discussed that all the information of the invariants under the marginal deformation is encoded in the index \cite{Kinney:2005ej}. Thus the computation of the finite $N$ index is of great interest and is  one of the central theme of this paper.

The index for the $\mathcal{N}=4$ SYM theory on $S^3 \times R$ is defined as a (twisted) Witten index
\begin{eqnarray}
\mathcal{I}(t,y,v,w) = \mathrm{Tr} (-1)^F e^{-\beta \Delta} t^{2(E+j_2)} y^{2j_1} v^{R_2} w^{R_3} \ ,
\end{eqnarray}
where $E$ is the energy (or conformal dimension via conformal mapping from $S^3 \times R$ to $R^4$), $j_1, j_2$ denote the spin quantum numbers associated with the rotation around $S^3$ whose isometry is $SU(2)_1\times SU(2)_2$, and $R_2, R_3$ denote residual R-charges in $SO(6)$ that commute with the regularization factor 
\begin{eqnarray}
\Delta = 2\{Q^{-\dagger}_1, Q^-_1\} = E-2j_2 - \frac{3}{2} r = E-2j_2 - 2\sum_{k=1}^3 \frac{4-k}{4} R_k \ .
\end{eqnarray}
Here $r$ denotes a particular $\mathcal{N}=1$ R-symmetry in the $SO(6)$ R-symmetry of $\mathcal{N}=4$ superconformal algebra while $R_k$ denotes three Cartan subgroups of $SO(6)$ in the $SU(4)$ notation. Due to the Bose-Fermi cancellation, the contribution to the index only comes from the states satisfying $\Delta = 0$, and as an immediate consequence, the index does not depend on $\beta$. We also note that if we omit the chemical potential related to $R_2$ and $R_3$, we can define the index for any $\mathcal{N}=1$ SCFTs by identifying $r$ with the $U(1)$ R-symmetry of the $\mathcal{N}=1$ superconformal algebra.

The index for a large class of SCFTs that can be flown to a free field theory by a marginal deformation can be computed by using a matrix model integral \cite{Sundborg:1999ue,Aharony:2003sx,Kinney:2005ej,Nakayama:2005mf}. Explicitly the index for the $\mathcal{N}=4$ SYM theory is computed as 

\begin{eqnarray}
\mathcal{I}(t,y,v,w) &\equiv& \mathrm{Tr} (-1)^F t^{2(E+j_2)} y^{2j_1} v^{R_2} w^{R_3}\cr
&=&\int [dU] e^{-S_{eff}[U]} \ , \label{matrix}
\end{eqnarray}
where the effective matrix action is given by
\begin{eqnarray}
-S_{eff}[U] = \cr
 \sum_{n>0} \frac{1}{n}\frac{t^{2n}(v^n+w^{-n}+w^{n}v^{-n})-t^{3n}(y^n+y^{-n})-t^{4n}(w^n+v^{-n}+v^nw^{-n})+2t^{6n}}{(1-y^{n}t^{3n})(1-y^{-n}t^{3n})}\chi_n(U) \ . 
\end{eqnarray}
For $SU(N)$ gauge group, the character for the adjoint representation is given by $\chi_n(U) = [ \mathrm{Tr}U^n\mathrm{Tr} U^{-n} - 1]$ and for  $U(N)$ gauge group it is given by $\chi_n(U) = \mathrm{Tr}U^n\mathrm{Tr} U^{-n}$. 
The integration is taken over the (special) unitary matrix $U$ with the invariant Haar measure $[dU]$. 

The exact integration is possible in $N\to \infty$ limit by using the saddle point approximation valid in the large $N$ matrix integral, which yields an elegant  expression for  $U(N)$ gauge group: 
\begin{eqnarray}
\mathcal{I}_{U(\infty)} = \prod_{n>0} \frac{(1-t^{3n} y^n)(1-t^{3n}y^{-n})}{(1-t^{2n}/w^{n})(1-t^{2n}w^n/v^n)(1-t^{2n} v^n)} \ . 
\end{eqnarray}

It is a challenging task to compute the finite $N$ index from the direct evaluation of the matrix integral \eqref{matrix}. For instance, in the $SU(2)$ case,\footnote{For simplicity, we have set $v=w=1$.}
one can rewrite the matrix integral as
\begin{eqnarray}\frac{2}{\pi}\int_0^{\pi} d\theta \sin^2\theta  \cr
\prod_{m>0}\left[(1-t^{3m}y^m e^{2 i\theta})(1-t^{3m}y^m e^{-2 i\theta})(1-t^{3m}y^m)\right] \times \cr
\times\prod_{m>0}\left[(1-t^{3m}y^{-m} e^{2 i\theta})(1-t^{3m}y^{-m} e^{-2 i\theta})(1-t^{3m}y^{-m}) \right]\times  \cr
\times \prod_{m,l\ge 0} \frac{(1-t^{3(m+l)+4}y^{-(m-l)} e^{2 i\theta})^3(1-t^{3(m+l)+4}y^{-(m-l)} e^{-2 i\theta})^3(1-t^{3(m+l)+4}y^{-(m-l)})^3}{(1-t^{3(m+l)+2}y^{-(m-l)} e^{2 i\theta})^3(1-t^{3(m+l)+2}y^{-(m-l)} e^{-2 i\theta})^3(1-t^{3(m+l)+2}y^{-(m-l)})^3} \ 
\end{eqnarray}
in the eigenvalue basis.
It is possible to rewrite the first two lines by using $\vartheta_{1}(t^{3}y^{\pm 1},e^{2i\theta})$, but it seems difficult to perform the integration in a closed form.

To facilitate the computation of the finite $N$ index, we can take several simplifying limits. For example, the half BPS limit has been studied in \cite{Kinney:2005ej}, where we take $t\to 0$ while keeping $t^2 v$ fixed. This limit enables us to compute the matrix integral exactly and even more we can write down the generating function of the finite $N$ index in a concise closed form:
\begin{eqnarray}
\sum_{N} p^N \mathcal{I}_{U(N)} &=& \prod_{n\ge0} \frac{1}{(1-p t^{2n}v^n)} \ . \label{gen}
\end{eqnarray}
Its large $N$ limit is obtained as 
\begin{eqnarray}
\mathcal{I}_{U(\infty)} = \lim_{p\to 1} (1-p)\sum_{N} p^N \mathcal{I}_{U(N)} = \prod_{n>1} \frac{1}{1-t^{2n}v^n} \ , \label{half}
\end{eqnarray}
which exactly coincides with the interacting half BPS partition function of the $\mathcal{N}=4$ SYM theory \cite{Kinney:2005ej}.\footnote{The appearance of the concise generating function is related the Plethystic Program studied in \cite{Benvenuti:2006qr,Feng:2007ur} although the definition of our index and BPS partition functions is different from their mesonic half BPS partition function.} The result is beautiful but the problem is that this kind limit is only possible for  theories with extended superconformal algebra. 

In this paper, we take yet another limit that makes the computation of the finite $N$ index feasible but does not require the extended superconformal algebra. Namely, we set $t\to 0$, $y \to {\infty}$ while keeping $t^{3}y$ fixed. In the large $N$ limit, we obtain
\begin{eqnarray}
\mathcal{I}_{U(\infty)} = \prod_{n>0} (1-t^{3n}y^n) \ ,
\end{eqnarray}
and we can regard it as a fermionic counterpart of the half BPS limit \eqref{half}.
Unlike the half BPS bound discussed above, this limit is not protected from the superconformal algebra, but as we will see later in section \ref{sec:3}, the limit naturally stems from the supergravity dual as the maximal angular momentum limit of the supergravity fields. In this limit, the explicit evaluation of the index and the BPS partition function is possible and we show that the index does {\it not} receive any finite $N$ corrections while the free BPS partition function does.\footnote{We emphasize that the index in a general parameter region {\it does} receive finite $N$ corrections.} From the supergravity viewpoint, the appearance of the limit is universal, independently of the internal Sasaki-Einstein manifold, and it is applicable to any SCFTs that possess a gravitational description. 

The organization of the paper is as follows. In section \ref{sec:2} we compute the index and the free BPS partition function for finite $N$ in the maximal angular momentum limit. In section \ref{sec:3} we present motivations to study the maximal angular momentum limit from the dual supergravity perspective. We also show that the known AdS black hole solutions do not contribute to the index in this limit. In section \ref{sec:4} we conclude with some discussions. In appendices, we summarize relevant superconformal algebra and its representation.

\sectiono{Finite $N$ Corrections in Maximal Angular Momentum Limit}\label{sec:2}
\subsection{Index}
As discussed in the introduction, we are interested in the maximal angular momentum limit of the finite $N$ index: $t\to 0$, $y \to {\infty}$ while keeping $t^{3}y$ fixed. Noting that the index has contributions only from the states that satisfy $\Delta = E-2j_2 - \frac{3}{2} r = 0$, we see that the limit picks up the states with $E = 3j_1 - j_2$ or equivalently $\frac{r}{2} = j_1-j_2$.\footnote{Here we have assumed that $E-3j_1+j_2 \ge 0$. The existence of this bound is deeply connected with the AdS-CFT duality as we will see.} Since the limit prevents us from taking too large $j_1$ quantum number with fixed $j_2$, we call it as the maximal angular momentum limit for the BPS operators.

In the free SYM theory, the index in this limit has a contribution from ${\lambda}_{+0}$ and its derivatives $\partial_{++}$. The bound $E \ge 3j_1 - j_2$ is certainly satisfied in the free SYM theory as can been seen from table \ref{tab:1}. The limit, however, is not guaranteed from the superconformal algebra. Indeed one could imagine a hypothetical operator that violates the bound such as the one that satisfies $E=\frac{3}{2}r$ (hence $j_2=0$) but has a large $j_1$, say $j_1 = \frac{1}{2}r+1$ violating the maximal angular momentum limit.\footnote{From the unitarity of the $SU(2,2)$ algebra (see appendix \ref{sec:A}), too large $j_1$ angular momentum is forbidden, but the maximal angular momentum bound is stronger than this unitarity bound.}  As we will discuss later, nevertheless, in the large $N$ dual supergravity viewpoint, the appearance of the bound is natural from the maximal angular momentum limit of the supergravity.
 
The limit facilitates the computation of the index for finite $N$. Let us illustrate how the matrix integral \eqref{matrix} reduces to a tractable form in the simplest nontrivial example of $SU(2)$: 
\begin{eqnarray}
\mathcal{I}_{SU(2)} &=& \frac{2}{\pi}\int_0^{\pi} d\theta \sin^2\theta 
\prod_{m>0}\left[(1-t^{3m}y^m e^{2 i\theta})(1-t^{3m}y^m e^{-2 i\theta})(1-t^{3m}y^m)\right]  \cr
&=& -\frac{1}{\pi}(t^3y)^{-\frac{1}{8}} \int_0^\pi d\theta \sin\theta \vartheta_1 (t^3y,e^{2i\theta}) \cr
&=& 1 
\end{eqnarray}
where the last equality comes from the integration over the theta function with characteristics ${\vartheta}_1(t^3y,e^{2i\theta})$. Therefore every terms cancel except for the identity operator.

Actually, this result seems universal: for any finite $N$, we have checked by using computer that $t\to 0$, $y \to {\infty}$ limit of the index for $SU(N)$ gauge group gives a trivial index. This is especially true in the large $N$ limit, where an explicit formula is available. In the large $N$ limit for $U(N)$ gauge group, the saddle point approximation of the matrix integral yields
\begin{eqnarray}
\mathcal{I}_{U(\infty)} = \prod_{n> 0} (1-t^{3n}y^n) \ , \label{inl}
\end{eqnarray}
while the contribution from the $U(1)$ decoupled degrees of freedom yields
\begin{eqnarray}
\mathcal{I}_{U(1)} &=& \exp\left[\sum_{n>0} \frac{1}{n}\frac{2t^{6n}-t^{3n}(y^n+y^{-n})+3t^{2n}-3t^{4n}}{(1-y^{n}t^{3n})(1-y^{-n}t^{3n})}\right] \ \cr
                   &\to& \prod_{n> 0} (1-t^{3n}y^n)  \ ,
\end{eqnarray}
in the maximal angular momentum limit.
As a consequence, only the free $U(1)$ part contributes to the index in this limit.\footnote{At this point, we note that the index does not depend on the coupling constant of the gauge theory, and the computation of the index from the supergravity, which we will not repeat here (see \cite{Kinney:2005ej,Nakayama:2005mf,Nakayama:2006ur} for various models), reveals an excellent agreement with the large $N$ result \eqref{inl}. }
We thus conclude that the index in this particular limit does not receive any finite $N$ corrections. 

\begin{table}[tb]
\begin{center}
\begin{tabular}{c|c|c|c}
 Letters        & $(-1)^F[E,j_1,j_2] $& $[R_1,R_2,R_3]$ & $r$\\
 \hline
 $X,Y,Z$&     $  [1,0,0] $         &$[0,1,0],[1,-1,1],[1,0,-1]$ & $\frac{2}{3}$ \\
 \hline
   $\psi^{X,Y,Z}_{0+}$&     $  -[\frac{3}{2},0,\frac{1}{2}] $         &$[1,-1,0],[0,1,-1],[0,0,1]$& $\frac{1}{3}$ \\
  $\lambda_{\pm 0}$&     $  -[\frac{3}{2},\pm\frac{1}{2},0] $ &$[1,0,0] $& $1$ \\
\hline
$F_{++}$&     $ [2,0,1] $ &$[0,0,0] $&$0$ \\
\hline
$\partial_{++}{\lambda}_{-0} + \partial_{-+}{\lambda}_{+0}=0  $&     $ - [\frac{5}{2},0,\frac{1}{2}] $ &$ [1,0,0] $& $1$ \\
\hline
$\partial_{\pm+} $&     $ [1,\pm \frac{1}{2},\frac{1}{2}] $ &$0 $& $0$ 
 \end{tabular}
\end{center}
\caption{List of the letters contributing to the index (hence $\Delta = 0$).}
\label{tab:1}
\end{table}%

\subsection{Free BPS partition function}
Here we would like to compute the BPS partition functions instead of the index in the same limit discussed above. Before the actual computation, we recall that the BPS partition function is not protected from marginal deformations of the SCFT. We concentrate on the free ($g_{YM}=0$) case in this subsection, and we deal with the interacting case in the next subsection.

The definition of the BPS partition function \cite{Kinney:2005ej} is
\begin{eqnarray}
\mathcal{Z}(t,y) = \mathrm{Tr}_{\Delta=0} t^{2(E+j_2)} y^{2j_1} \ .
\end{eqnarray}
The trace is taken over the states that satisfy $\Delta = E-2j_2-\frac{3}{2}r = 0$.\footnote{Since there is no need for the chemical potential to commute with the supercharge, one could (but we will not) introduce an additional chemical potential, e.g. $x^{2E}$ other than $v^{R_2}w^{R_3}$.}

The computation of the BPS partition function in the zero coupling limit of the  $\mathcal{N}=4$ SYM theory can be performed in the similar way as before by the matrix integral. The only difference is that the effective matrix action has an additional sign factor in front of the fermionic contribution:
\begin{eqnarray}
-S_{eff}[U] = \sum_{n>0} \frac{1}{n}\frac{3t^{2n}-(-1)^{n+1}t^{3n}(y^n+y^{-n})-(-1)^{n+1}3t^{4n} + 2t^{6n}}{(1-y^{n}t^{3n})(1-y^{-n}t^{3n})}\chi_n(U) \ .
\end{eqnarray}

Let us compute the matrix integral explicitly in the maximal angular momentum limit. We take $y\to \infty$, $t \to 0$ limit and set $q=t^3 y$ for notational simplicity in the following.
In the $SU(2)$ case, we obtain the following result:
\begin{eqnarray}
\mathcal{Z}_{SU(2)} &=& \frac{2}{\pi} \int_{0}^{\infty} \sin^2\theta \prod_{n>1} (1+t^{3n}y^n e^{2i\theta})(1+t^{3n}y^n e^{-2i\theta})(1+t^{3n}y^n) \cr
&=& 1 + 2q^3 + 2q^4 + 4q^5 + 6q^6 + 8q^7 + \cdots .
\end{eqnarray}
Note that we have obtained nontrivial series in contrast to the index.

Similarly, for $SU(3)$ we obtain 
\begin{eqnarray}
\mathcal{Z}_{SU(3)} &=& 1 + 2q^3 + 2q^4 + 6q^5 + 12q^6 + 20q^7 + \cdots .
\end{eqnarray}
It would be interesting to derive a generating function of the finite $N$ free BPS partition functions in the maximal angular momentum limit as in \eqref{gen}.

The $SU(N)$ BPS partition function in the large $N$ limit with $g_{YM}=0$ can be computed by using the saddle point approximation,
 which results in 
\begin{eqnarray}
\mathcal{Z}_{SU(\infty)} &=& \prod_{n>0} \frac{(1-q^n)}{(1+(-1)^nq^n-q^n)(1+q^n)} \cr
&=& 1 + 2q^3 + 2 q^4 + 6q^5 + 12 q^6 + 22 q^7 + \cdots \ . \label{flN}
\end{eqnarray}

In all cases treated above, we can multiply the free decoupled $U(1)$ contribution
\begin{eqnarray}
\mathcal{Z}_{U(1)} = \prod_{n>0} (1+q^n) \ .
\end{eqnarray}
to obtain the BPS partition function for $U(N)$ gauge theories instead of $SU(N)$ gauge theories.

It is interesting to note that the supergravity BPS partition function coincides with the free $U(1)$ BPS partition function:
\begin{eqnarray}
\mathcal{Z}_{gravity}= \mathcal{Z}_{U(1)} = \prod_{n>0} (1+q^n) \ . \label{gz}
\end{eqnarray}
The computation of the BPS partition function from the supergravity dual is straightforward (see \cite{Kinney:2005ej}) and we will not repeat it here. The appearance of the fermionic partition function for a harmonic oscillator is due to the fact that the fundamental building block is given by the free fermion living on the boundary of the AdS space.
The disagreement between \eqref{flN} and \eqref{gz} is not unexpected because the BPS partition function is not protected under the marginal deformation of the SCFT.

\subsection{Weakly interacting BPS partition function}
We have seen that the free BPS partition function in the large $N$ limit does not coincide with the BPS partition function computed from the gravity dual. It is known that the BPS partition function jumps once we turn on the non-zero Yang-Mills coupling constant. Furthermore, there is a widely spread conjecture for the $\mathcal{N}=4$ SYM theory that there does not exist any additional quantitative change of the BPS partition function throughout every non-zero coupling constant to the supergravity limit. 

In our particular maximal angular momentum limit ($y\to \infty$, and $t \to 0$ while keeping $q=t^3 y$ fixed), we have seen that in the strongly coupled limit, the BPS partition function is same as the BPS partition function of the free decoupled $U(1)$ contribution.  This naturally leads us to the conjecture that all the non-Abelian parts become non-BPS and do not contribute to the interacting cohomology\footnote{From the natural identification of $Q^-_{1}$ with exterior derivative, we use the standard terminology of cohomology to describe BPS states.} in contrast to the free cohomology discussed in the last subsection. 
 
Indeed, one can show this statement in the weakly coupled (but not free) $\mathcal{N}=4$ SYM theory as follows. We first note that the derivatives $\partial_{++}$ appearing in the operators should be replaced by its covariantized counterpart $D_{++}$. We then notice the commutation relation
\begin{eqnarray}
[Q_{-},D_{++}]\cdot O = -2[{\lambda}_{+0},O] 
\end{eqnarray}
for any operator $O$ in the adjoint representation of $SU(N)$. 
In particular, we have the following relation (see e.g. \cite{Berkooz:2006wc})
\begin{eqnarray}
\left\{Q_{-}, \underbrace{D_{++}\cdots D_{++}}_{I \ \text{times}} {\lambda}_{+0} \right\} = -2\sum_{m=1}^I \begin{pmatrix} I \cr m 
\end{pmatrix} \left\{\underbrace{D_{++}\cdots D_{++}}_{(m-1) \ \text{times}} {\lambda}_{+0}, \underbrace{D_{++}\cdots D_{++}}_{(I-m) \ \text{times}} \lambda_{+0} \right\} \ .\label{interacting}
\end{eqnarray}
From these relations, the non-Abelian part of the BPS operators in the free cohomology combine themselves to make a long non-BPS multiplet and do not contribute to the BPS partition functions in the weakly coupled cohomology. For instance, let us take a look at the first nontrivial example at level $q^3$. The free BPS partition functions have contributions from four operators: $\mathrm{Tr} (\partial_{++}^2 \lambda_{+0})$, $\mathrm{Tr} (\partial_{++}{\lambda}_{+0})\mathrm{Tr}({\lambda}_{+0})$, $\mathrm{Tr} (\partial_{++}{\lambda}_{+0}{\lambda}_{+0})$, $\mathrm{Tr} ({\lambda}_{+0}{\lambda}_{+0}{\lambda}_{+0})$, the first two of which are $U(1)$ contribution and survive in the interacting cohomology. In the free cohomology $\mathrm{Tr} (\partial_{++}{\lambda}_{+0}{\lambda}_{+0})$ belongs to a  LH-semi-long multiplet while $\mathrm{Tr} ({\lambda}_{+0}{\lambda}_{+0}{\lambda}_{+0})$ belongs to a chiral LH-multiplet independently. However, in the weakly interacting cohomology, due to the relation \eqref{interacting}, they combine themselves to make a long multiplet. A long multiplet is not protected from the renormalization so that it will eventually violate the BPS condition and do not contribute to the BPS partition function in the interacting cohomology. In this way, in the weakly interacting cohomology of the $\mathcal{N}=4$ SYM, only the free decoupled $U(1)$ part contributes and it reproduces the strongly coupled gravity limit \eqref{gz}.

\sectiono{Maximal Angular Momentum Limit from Gravity}\label{sec:3}
\subsection{Supergravity limit}
So far we have studied the maximal angular momentum limit without explicitly stating any physical motivations. In this section, we present the motivation to take the limit from the dual supergravity viewpoint.

In the supergravity description of the gauge invariant operators, we indeed have a natural bound for the angular momentum because the underlying supergravity field has a maximal spin two from the type IIB supergravity. This is nothing but the well-known statement that one cannot construct nontrivially interacting field theory from finitely many massless higher spin fields than two (see e.g. \cite{Sorokin:2004ie} and references therein). 
 After the Kaluza-Klein decomposition of the type IIB supergravity on $AdS_5\times X_5$, where $X_5$ denotes any Sasaki-Einstein manifold, we will obtain the maximal angular momentum limit introduced earlier in this paper.

We first note that the BPS state contributing to the index falls into two distinct representations of the $SU(2,2|1)$ superconformal algebra as reviewed in appendix \ref{sec:A}. We further demand from the supergravity dual description that the angular momentum appearing in the highest representation of the $SU(2,2)$ conformal algebra is bounded above by $j_1, j_2 \le 1$.\footnote{$SU(2,2)$ descendants of these operators will generate higher spin operators, but it is easy to see that acting $P_{-+}$ or $P_{++}$ does not violate the maximal angular momentum condition. Indeed, acting $P_{++}$ produces higher spin states that continue to saturate the bound.} 

Let us begin with the chiral LH-multiplet. The only possibility here is to take $j_1 = 0$ or $j_1=\frac{1}{2}$. When $j_1= 0$, the saturation of the angular momentum limit $E \ge 3j_1 - j_2$ is obvious. When $j_1 = \frac{1}{2}$, the saturation of the angular momentum limit is guaranteed from the unitarity: $E \ge 3j_1 - j_2 = \frac{3}{2} = j_1 + 1$. The saturation of the unitarity bound $E \ge j_1 + 1 $ (for $j_2=0$) corresponds to a free decoupled degree of freedom in consistent with the analysis in the previous sections.

The other case is the LH-semi-long multiplet. Again the possibilities of the angular momenta are limited: $(j_1,j_2) = (0,0),(\frac{1}{2},0),(0,\frac{1}{2}),(\frac{1}{2},\frac{1}{2})$ for the $SU(2,2|1)$ primary, so the BPS operators contributing to the index necessarily have $(j_1,j_2) = (0,\frac{1}{2}),(\frac{1}{2},\frac{1}{2}),(0,1),(\frac{1}{2},1)$  The maximal angular momentum bound is obviously satisfied from the unitarity because $ E \ge 1 \ge 3j_1-j_2$. 

In this way, we have shown that the maximal angular momentum limit is naturally predicted from the supergravity dual description of the gauge theory. In the large $N$ strong coupling limit, all the BPS states should satisfy this bound in any field theories that have a supergravity dual. We stress that this is nontrivial because there is no apparent derivation of this bound from the interacting field theory.\footnote{For a free SCFT, the bound can be directly checked from the bare quantum numbers as we did in table \ref{tab:1} for the $\mathcal{N}=4$ SYM theory.} In addition, the gravity description suggests that the saturation of the bound is achieved by the free decoupled degrees of freedom. In the gauge theory, they are typically decoupled $U(1)$ degrees of freedom. This is consistent with our previous results: the index and the weakly coupled BPS partition function do not receive finite $N$ corrections in the maximal angular momentum limit. It also explains the emergence of the finite $N$ corrections in the free BPS partition functions, and its disappearance after taking account of the weak but non-zero gauge coupling constant. No matter how tiny the gauge coupling constant is, it will make the theory non-free for the non-Abelian part, and the saturation of the maximal angular momentum bound would be violated.

\subsection{Contribution from AdS black holes}
To understand finite $N$ contributions to the index and BPS partition functions from the gravity side, the contributions from the AdS black hole solutions would be important in the high energy limit. If we restrict ourselves to the case of equal R-charges of $SO(2)\times SO(2) \times SO(2) \in SO(6)$,\footnote{The R-charge notation here is different from before: $q_{1}^{SO(6)} = \frac{R_1^{SU(4)}}{2} + R_2^{SU(4)} + \frac{R_3^{SU(4)}}{2} \ , \  q_{2}^{SO(6)} = \frac{R_1^{SU(4)}}{2} + \frac{R_3^{SU(4)}}{2} \ , \  q_{3}^{SO(6)} = \frac{R_1^{SU(4)}}{2} -\frac{R_3^{SU(4)}}{2}$. Since $\lambda_{+0}$ possesses equal $q_i^{SO(6)}$ R-charges in this $SO(6)$ convention, this restriction is relevant to us.} the BPS black holes obtained in \cite{Gutowski:2004ez,Gutowski:2004yv,Chong:2005hr,Kunduri:2006ek} possess the following characteristic charges
\begin{eqnarray}
J_2 + J_1 &=& N^2 \frac{(a+b)(2a+b+ab)}{4(1-a^2)(1-b)} \cr
J_2 - J_1 &=& N^2 \frac{(a+b)(a+2b+ab)}{4(1-a)(1-b^2)} \cr
R &=& N^2 \frac{a+b}{(1-a)(1-b)} \cr
E &=& 2J_2 + \frac{3}{2}R = N^2\frac{(a+b)[(1-a)(1-b)+(1+a)(a+b)(2-a-b)]}{4(1-a)^2(1-b)^2} \  \label{parameter}
\end{eqnarray}
parametrized by two independent real numbers $a$ and $b$.
To avoid the existence of a naked closed time-like curve, the parameter region is restricted to $|a|,|b| <1$ and $ a+b+ab>0$. The last equality in \eqref{parameter} corresponds to the saturation of the BPS bound.

In general, for finite $N$, these black holes will contribute to the index and the BPS partition functions as well (see, however, the discussion given in \cite{Kinney:2005ej} for a possibiliy of null contributions to the index from the black holes).
Now let us consider the particular limit we have been interested in: $y\to \infty$, and $t \to 0$ while keeping $q=t^3 y$ fixed. In this limit, only the states satisfying $J_1-J_2 = \frac{R}{2}$ (together with the BPS bound $E=2J_2 + \frac{3}{2}R$) will contribute. However it is easy to see that there do no exist any black holes that satisfy this condition because $J_2-J_1 > 0 > -\frac{R}{2}$ for $R>0$. This is consistent with the fact that there is no finite $N$ correction to the index and the weakly interacting BPS partition function in this limit. 

Several remarks are in order
\begin{itemize}
	\item It has not been established that \eqref{parameter} completes the BPS black holes that are asymptotically AdS (but see \cite{Kunduri:2006uh} for a support of this claim). Thus one cannot say that the black holes really do not contribute to the index or the BPS partition function in this limit. However, our result that the index and the weakly interacting BPS partition function do not depend on $N$ strongly suggests that the index and the interacting BPS partition function in this maximal angular momentum limit receive contributions only coming from the free decoupled $U(1)$ degrees of freedom. 
	\item In the literature \cite{Berkooz:2006wc}, it has been discussed that there is a connection between the BPS black holes satisfying $J_1 \sim J_2 $ and the operators made from ${\lambda}_{+0}$ and derivatives $D_{++}$. Strictly speaking, these operators have a charge $J_1-J_2 = \frac{R}{2}$ and cannot occur from the black holes as discussed above. However, in the large $N$ and $J\gg |R|\gg 1$ limit, the quantitative agreement of the entropy for such operators (up to a numerical factor of order one) has been observed.
	
\end{itemize}

The microscopic description of the BPS black holes discussed here is supposed to be related to the $1/16$-BPS (dual) giant gravitons in the AdS space (see \cite{Kim:2006he} for a recent discussion). So far, the complete classification of the solutions and their quantization are unavailable unlike the $1/8$-BPS sector. It would be very interesting to study the contributions to the index and the BPS partition functions from these objects. In particular, the role of the maximal angular momentum limit in their quantization is an intriguing subject to study.

\sectiono{Discussion and Conclusion}\label{sec:4}
In this paper, we have studied the finite $N$ index and the BPS partition functions in the newly proposed maximal angular momentum limit. In this limit, the explicit evaluation of the matrix integral is feasible and we have shown that the index and the weakly interacting BPS partition function do not receive any finite $N$ corrections while the free BPS partition function does. In a general parameter region, even the index receives finite $N$ corrections, and it would be important to reveal the effect of finite $N$ corrections in departure from the maximal angular momentum limit.

The physical nature of the  maximal angular momentum limit needs further studying. In the supergravity limit, we understand it as the maximal angular limit from the Kaluza-Klein decomposed type IIB supergravity fields. Do string states satisfy the maximal angular limit? They certainly do not contribute to the index because the index is independent of the coupling constant (and hence any $\alpha'$ corrections), but the contribution to the BPS partition function is an interesting problem. It would be also interesting to study the non-BPS states near the maximal angular momentum limit, which might give rise to yet another solvable limit of the AdS-CFT correspondence.

As discussed in the introduction, any finite $N$ enumeration program in the SCFTs with gravity dual would be of significance in order to understand the quantum nature of the nonperturbative string theory. In this respect, the connection between our results and other enumeration programs for SCFTs \cite{Martelli:2006yb,Alvarez-Gaume:2006jg,Biswas:2006tj,Mandal:2006tk,Sinha:2006ac,Hikida:2006qb,Bianchi:2006ti,Butti:2006au,Hanany:2006uc} is of great interest. Especially, the Plethystic Program \cite{Benvenuti:2006qr,Feng:2007ur} is fascinating in the sense that we can write down the generating function of finite $N$ mesonic partition functions from the data of the geometry. We expect that the finite $N$ index, which we emphasize is the only invariant of the SCFT under the marginal deformation, should also have a deep connection to the geometry. We hope we could answer these problems in the near future.

\section*{Acknowledgements}
The author would like to acknowledge their hospitality in his stay at Perimeter Institute, where the initial stage of this work was done.
This research is supported in part by JSPS Research Fellowships
for Young Scientists.

\appendix\sectiono{Superconformal algebra $SU(2,2|1)$}\label{sec:0}
Relevant (anti-)commutation relations of the superconformal algebra $SU(2,2|1)$  for our discussion are
\begin{align}
[(J_2)^\alpha_\beta, (J_2)^{\gamma}_{\delta}] &= \delta^{\gamma}_\beta (J_2)^\alpha_\delta - \delta^{\alpha}_\delta (J_2)^\gamma_\beta   \ ,\cr
[H,P^{\dot{\alpha}{\beta}}] &= P^{\dot{\alpha}{\beta}}  \ ,\cr
[H,Q^{\gamma}] &= \frac{1}{2} Q^\gamma \ ,\cr
[r,Q^{\gamma}] &= Q^\gamma  \ ,\cr
[r,P^{\dot{\alpha}{\beta}}] &= 0 \ ,\cr
\{ S_{\alpha},Q^\beta \} &= (J_2)^\beta_{\alpha} +\delta^\beta_{\alpha} \left(\frac{H}{2} - \frac{3}{4} r\right)  \ ,
\end{align}
where upper dotted spinor indices denote $SU(2)_1$ fundamental representation and un-dotted spinor indices denote $SU(2)_2$ fundamental representation.

\sectiono{Representation of $SU(2,2|1)$}\label{sec:A}
In this appendix, we summarize the classification of unitary representations of the superconformal algebra $SU(2,2|1)$ \cite{Flato:1983te} (see appendix B of \cite{Freedman:1999gp} for a review). The representation of the superconformal algebra $SU(2,2|1)$ is naturally decomposed into a highest weight representation of its bosonic subalgebra $SU(2,2) \simeq SO(4,2)$ (conformal algebra) and $U(1)_r$ (R-symmetry). Accordingly, the highest weight representation of $SU(2,2|1)$ is classified by four quantum numbers $D(E_0,j_1,j_2;r)$. 

The unitary representation of $SU(2,2|1)$ with generic quantum number $E_0,j_1,j_2,r$ consists of a long multiplet with $16$ $SU(2,2)$ highest weight representation. However, if one of the following combination of the quantum number vanishes, the representation becomes short.
\begin{align}
n_1^{(+)} &= N(E_0 + 2j_1 + \frac{3}{2}r)  \cr
n_1^{(-)} &= N(E_0 - 2j_1 + \frac{3}{2}r-2)  \cr
n_2^{(+)} &= N(E_0 + 2j_2 - \frac{3}{2}r)  \cr
n_2^{(-)} &= N(E_0 - 2j_2 - \frac{3}{2}r-2)  \cr
s_1 &= N(j_1) ,   \ \ \mathrm{and} \ \ \ \ s_2=N(j_2) \ , \label{short}
\end{align}
where, following \cite{Freedman:1999gp}, we have introduced the false-function: $N(x) =1$ for $x\neq 0$ and $N(0) =0$.
 The multiplet shortening structure is summarized in table \ref{table1}, which is borrowed from \cite{Freedman:1999gp} with corrections.

What is relevant for computing the index for SCFT on $S^3 \times R$ is which (short) multiplet contributes to the index. Recalling that the states contributing  to the index  satisfy $\Delta = E - 2j_2 -\frac{2}{3} r = 0$, we find {\it chiral LH-multiplets} with quantum number $D(E_0,j,0;r)$ with $r=\frac{2}{3}E_0$ and {\it LH-semi-long multiplet} $D(E_0,j_1,j_2;r)$ with $r=\frac{2}{3}(E_0-2j_2-2)$ (together with their specific $SU(2,2)$ descendants) will contribute to the index. The structure of these short multiplets contributing to the index are presented in table \ref{table2} and \ref{table3}.

Let us now consider the contribution to the index from $SU(2,2)$ descendants obtained by acting $P_{\dot{\alpha}{\alpha}}$ to $SU(2,2)$ primaries. It is clear only $P_{\pm+}$ will generate descendants contributing to the index by acting on $SU(2,2)$ primaries with vanishing $\Delta$. However, there is a possibility that $SU(2,2)$ descendants show  a multiplet shortening \cite{Mack:1975je}. In this case, we have to subtract contributions from their null vectors. This happens when 
\begin{align}
(a) \ \ \ \ \ \ j_1j_2 &\neq 0 \ ,  \ \ \ E_0 = 2 + j_1 + j_2  \cr
(b) \ \  \ \ \ \ j_1j_2 &= 0    \ , \ \ \ E_0 = 1+j \cr
(c) \ \ j_1=j_2 &= 0 \ , \ \ \ E_0 = 0 \ .
\end{align}
In the main text, we have encountered the condition $(b)$ interpreted as the Dirac equation of the gaugino. 

\begin{table}[bp]
\begin{center}
\begin{tabular}{c|c|c} 
 Level & $SU(2,2)$ representation & Multiplicity \\\hline
 $0$ & $D(E_0,j_1,j_2;r)$ & 1  \\\hline
 $1$&$D(E_0+\frac{1}{2},j_1+\frac{1}{2},j_2;r-1)$  & $n_1^{(+)}$ \\
 & $D(E_0+\frac{1}{2},j_1-\frac{1}{2},j_2;r-1)$ &  $s_1n_1^{(-)}$\\
 & $D(E_0+\frac{1}{2},j_1,j_2-\frac{1}{2};r+1)$ & $s_2n_2^{(-)}$ \\
 & $D(E_0+\frac{1}{2},j_1,j_2+\frac{1}{2};r+1)$ & $n_2^{(+)}$ \\\hline
 $2$ & $D(E_0+1,j_1,j_2;r-2)$   & $n_1^{(+)}n_1^{(-)}$   \\
 &$D(E_0+1,j_1+\frac{1}{2},j_2+\frac{1}{2};r)$  &$n_1^{(+)}n_2^{(+)}$  \\
 &$D(E_0+1,j_1+\frac{1}{2},j_2-\frac{1}{2};r)$  &$s_2n_1^{(+)}n_2^{(-)}$ \\
 &$D(E_0+1,j_1-\frac{1}{2},j_2+\frac{1}{2};r)$  &$s_1n_1^{(-)}n_2^{(+)}$  \\
 & $D(E_0+1,j_1-\frac{1}{2},j_2-\frac{1}{2};r)$ &$s_1s_2n_1^{(-)}n_2^{(-)}$   \\
 & $D(E_0+1,j_1,j_2;r+2)$ &$ n_2^{(-)}n_2^{(+)}$ \\\hline
$3$ &$D(E_0+\frac{3}{2},j_1,j_2+\frac{1}{2};r-1)$  & $n_1^{(+)} n_1^{(-)} n_2^{(+)}$ \\
 & $D(E_0+\frac{3}{2},j_1,j_2-\frac{1}{2};r-1)$ & $s_2n_1^{(+)}n_1^{(-)}n_2^{(-)}$ \\
 & $D(E_0+\frac{3}{2},j_1-\frac{1}{2},j_2;r+1)$ &$s_1n_1^{(-)}n_2^{(+)}n_2^{(-)}$   \\
 & $D(E_0+\frac{3}{2},j_1+\frac{1}{2},j_2;r+1)$ & $n_1^{(+)}n_2^{(+)}n_2^{(-)}$  \\\hline
$4$ & $D(E_0+2,j_1,j_2;r)$ &$n_1^{(+)}n_1^{(-)} n_2^{(+)}n_2^{(-)}$   \\\hline
\end{tabular} 
\end{center}
\caption{A long multiplet of $SU(2,2|1)$ contains $16$ highest weight representations of $SU(2,2)$. When the unitarity bound is saturated or $j_1j_2=0$, multiplet shortening occurs as shown in the table.}
\label{table1}
\end{table}

\begin{table}[bp]
\begin{center}
\begin{tabular}{c|ccc} 
$E\backslash R$ & $r$ & $r-1$ & $r-2$ \\\hline
$E_0$ & $\diamondsuit (j,0)$  &  &  \\
$E_0 +\frac{1}{2}$ &  & $(j+\frac{1}{2},0)\oplus (j-\frac{1}{2},0)$ &  \\
$E_0 +1$ &  &  & $(j,0)$ \\\hline
\end{tabular}
\end{center}
\caption{Chiral LH-multiplets $D(E_0,j,0;r)$ with $r=\frac{2}{3}E_0$. The representation with $\diamondsuit$ (in the top component of the table) contributes to the index. When $j=0$, further shortening occurs.}
\label{table2}
\end{table}

\begin{table}[bp]
\begin{center}
\begin{tabular}{c|cccc} 
 $E\backslash R$ & $r+1$ & $r$ & $r-1$& $r-2$\\\hline
 $E_0$&  & $(j_1,j_2)$ & & \\
$E_0+\frac{1}{2}$ & $\diamondsuit (j_1,j_2+\frac{1}{2})$ &  & $(j_1+\frac{1}{2},j_2)$ & \\
 &  &  &$(j_1-\frac{1}{2},j_2)$ & \\
 $E_0+1$ &  & $(j_1+\frac{1}{2},j_2+\frac{1}{2})$ & &$(j_1,j_2)$ \\
&  &$(j_1-\frac{1}{2},j_2+\frac{1}{2})$  & & \\
$E_0+\frac{3}{2}$ &  &  &$(j_1,j_2+\frac{1}{2})$ & \\\hline
\end{tabular}
\end{center}
\caption{LH-semi-long multiplet $D(E_0,j_1,j_2;r)$ with $r=\frac{2}{3}(E_0-2j_2-2)$.} The representation with $\diamondsuit$ (in the level one of the table) contributes to the index. When $j_1 =0$, further shortening occurs.
\label{table3}
\end{table}

\newpage
\bibliographystyle{utcaps}
\bibliography{angular}

\end{document}